\documentstyle{article}
\begin{document}
\begin{titlepage}
\centerline{\Large\bf Anyon Mean Field as an Exact  Limit of a Gauge Theory}
\centerline{Roberto Iengo}
\centerline{International School for Advanced Studies, SISSA and}
\centerline{Istituto Nazionale di Fisica
Nucleare, INFN, Sezione di Trieste, Trieste, Italy}

\begin{center}
\today
\end{center}
\begin{abstract}
We summarize a study of
an Abelian gauge theory in 2+1 dimensions, the gauge field
being coupled to nonrelativistic Fermions. The Action for the gauge
field is a combination of the Maxwell term and a Chern-Simons (CS)
term. We study the limit of vanishing Fermions' charge, keeping fixed
the gauge field mass induced by the CS term. By considering a closed
surface, in particular a torus to keep translational invariance, we show
that
the Fermions do not decouple completely from the gauge
field, and in fact they behave as Anyons treated in the
-translationally invariant formulation of the- Mean Field
approximation. We describe the exact solution of this limiting case.
\end{abstract}
\end{titlepage}
\def\pa{\partial}

This lecture is based on ref.\cite{MEAN} and on previous work done on
the problem of a translational invariant formulation of the Anyon mean
field theory and its properties with respect to superfluidity and possibly
superconductivity (for a general review of the previous work see
ref.\cite{Rep}). Actually, to discuss those issues in a proper way it is
necessary to introduce a suitable mathematical formalism: for instance the
proper definition of the translation operators requires the specification of
the boundary conditions, which turns out to give rise to a nontrivial
mathematical-physics problem of a field theory on a non-simply connected
surface, in our case a torus.

However, we have chosen here to emphasize the physical
approach to the problem and the very well defined physical results. In
fact we realized in ref\cite{MEAN} that the Anyon mean field problem can be
formulated as a limiting case, or in other words a critical case, of a more
general interesting field theory that, in the critical limit, can be
solved exactly. We decided then to skip the various mathematical-physics
technicalities and concentrate on the physics of the problem. It must be
said however that the more technical ingredients are absolutely essential
for the correct solution of the problem. For instance, the precise
determination of the spectrum can only be obtained after implementing the
modular invariance. A thorough discussion of these points does not usually
appear in the existing literature, and we are not aware of other papers
doing it (i.e. a precise statement of the boundary conditions and
their implications for the spectrum)
but ref.\cite{Rep} and our previous work quoted there.

We begin by considering a gauge theory in two space and one time
dimensions, the gauge field being coupled to nonrelativistic
Fermions. The Action for the gauge field is a combination of both the
standard Maxwell term $ -1/4e^2\int F^2$ and a Chern-Simons (CS) term
 $k/2\pi \int AF$. One can assume that this theory comes out as an
effective theory from some underlying microscopic dynamics in a
context of a condensed matter problem, like the Hubbard
model, see for instance \cite{BEMS} and references therein.
One expects both terms on general grounds. This (2+1)dimensional
gauge theory has been discussed in different contexts in various papers,
in particular in \cite{DJT}, \cite{SS}, \cite{Po1}.

The nonrelativistic particles are coupled to the gauge field
through the covariant derivative $D=\partial -iA$. We can rescale the
gauge field $A\to eA$ to get the covariant derivative to be
$D=\partial -ieA$ where we see that the parameter $e$ has the meaning
of the electric charge. One should of course remember that $e^2$ has
the dimensions of a mass because the dimensionality of the spacetime
is 3. The CS term can then be rewritten as ${\mu\over 2} \int AF$ where
$\mu =ke^2/2\pi$ and it is known that this term gives the mass
$\mu$ to the "photon", i.e. the quantum of the $A$ field becomes
massive \cite{DJT}.
It is also known that the limit $e\to \infty$ (at fixed $k$),
when the CS
term dominates over the Maxwell term, reproduces the CS theory where
there is no propagating particle associated to the $A$ field and the
effect of the gauge field is to introduce a statistics flip of the
particles to which it is coupled. And, in fact, one recovers in this
limit \cite{ILL}
the CS description of a system of Anyons, which are particles of
fractional statistics living on a two dimensional surface (for a general
review see \cite{WF} and references therein, and also \cite{Rep}
on more specific aspects of the Anyons' nonrelativistic quantum mechanics).

Here we consider instead the opposite limit $e\to 0$, while
keeping the mass $\mu$ finite. At first sight this limit looks trivial
because one expects the particles to decouple from the gauge field.
Moreover no possible infrared divergence effects should spoil this
naive expectation due to the finiteness of $\mu$.

But the problem is made interesting by specifying the boundary
conditions taking the two dimensional space, where the theory lives,
to be a closed surface. In particular we will take it to be a torus
because of the special interest in the aspects related to
translational invariance, the torus being the unique Riemann surface
which is invariant under translations. As the physical results are
independent of the surface's size, eventually we can take the
thermodynamical limit where the size of the torus goes to infinity at fixed
Fermions' density.

Since in our
case the nonrelativistic particles are $N$  identical Fermions,
then the total charge seen by the gauge field will be $Ne$.
It is well known that electromagnetism does not make sense on a
closed surface if the total charge is different from zero. However
our theory still exists, due to the CS term, provided the quantization
condition $N/k=integer$ is satisfied. One gets in fact a situation
which reminds the problem of charged particles in a magnetic field due
to monopoles, supposedly put in the region of a threedimensional space
bounded by the closed surface; the total magnetic flux through the surface
is different from zero but must obey a quantization condition. This
can be easily seen by considering the equation for $A_0$, in the
Coulomb gauge $\partial_i A_i=0$ ($i=x,y$):
$$
(\partial_x^2 +\partial_y^2 )A_0= e\rho +\mu F
$$
where $\rho$ is the particles' density and $F\equiv F_{xy}$. By
integrating both sides over the surface we get the relation
$$
\int Fd^2x=-{e\over \mu}\int \rho d^2x=-{2\pi \over ek}N
$$
This relation is a constraint on $F$, or in other words the theory
makes sense only in the Hilbert space where this relation holds.
Moreover we know that the total flux of $F$ can be different from
zero only when the Dirac quantization condition is satisfied,
namely$^{\cite{Ho1},\cite{Po3},\cite{IL1},\cite{IL2}}$
$$
\int Fd^2x= -{2\pi \over e}n
$$
giving $N/k=n=integer$. We can \cite{IL1,IL2} then
split $A_i$ into two pieces: $A_i=A_i^b+\tilde A_i$, where the
"background" part is fixed by the requirement
$$
F^b=-{2\pi \over ek}\rho_0,
$$
where $\rho_0$ denotes the average particles density.
There is no constraint
on $\tilde A_i$, which then represents the quantum fluctuating
(massive) degrees of freedom of the gauge field.
Actually this relation does not fix completely $A_i^b$, because
topological modes ("flat connections") are still undetermined:
they can be interpreted as Wilson loops of the gauge field along the
handles of the torus.

The explicit form of the gauge field is thus
(see ref. \cite{ILL} for a more
detailed discussion)
$$
A_i={a_i\over \nu e} +{\pi \over e\nu k}\epsilon_{ij}x_jN+\tilde A_i
$$
where the $a_i$ are the topological components (zero modes or "flat
connections")
of the gauge field, the
second term represents the background $A_i^b$ and the transverse
$\tilde A_i$ describes
the fluctuating part of the gauge field. $N$ is the total number of
Fermions, which is constant in our nonrelativistic problem.
Notice that $A_i$ cannot be globally defined, since its flux through the
surface is different from zero. This is seen here in the fact that $A_i$
is not periodic for $x_1\to x_1+m,~~x_2\to x_2+n\nu$ (the torus is
described as a rectangle by identifying opposite edges; the edges' sizes
are conventionally $1$ and $\nu$).
We thus only require
periodicity up to a gauge transformation (the fluctuating part $\tilde A$
is instead strictly periodic).
The independent variables of our quantum problem are the particles'
coordinates and the gauge field degrees of freedom $a_i$ and $\tilde A_i$ .
We use a
first quantization formalism and
the Hilbert space is thus made out of the  functionals of
$x_i^l, a_i$ and $\tilde A_i$, $\psi(\vec x^l,\vec a,\tilde A_i)$, where
the $x^l$ are the coordinates of the Fermions and $l$ runs from $1$ to $N$,
the (fixed) number of these particles. The wave function will be periodic up to
the above said gauge transformation.

In the limit $e\to 0$ the fluctuating part $\tilde A_i$ decouples
from the particles but the background does not since
$e\int F^b=-2\pi N/k$. In fact, in the limit $e\to 0$ we recover a
sort of a "mean field" situation, where the particles interact with a
constant magnetic-like field orthogonal to the surface and proportional
to the particles' density. The proportionality constant corresponds
precisely to the mean field approximation of the theory of Anyons,
also called sometimes
$^{\cite{La4},\cite{FHL1},\cite{FHL2},\cite{HC},\cite{CWWH}}$
"random phase approximation".

We remind \cite{WF} that the statistical parameter of these Anyons
is $\theta =\pi (1-1/k)$, where $\theta_0 =\pi$ is due to the fact that we
consider Fermions and $\Delta \theta =-\pi /k$ is the contribution from the
statistical field. The possibility that the limit $e\to 0$ could be related
to the mean field theory of Anyons was anticipated in ref.\cite{WF}. Notice
however that many important aspects of our picture, for instance keeping
$\mu$ finite and the role of the boundary conditions, are different from
the possible scenario considered in \cite{WF}.

This mean field theory looks very much like the problem of the
Landau levels, i.e. charged particles on a surface interacting with a
constant magnetic field orthogonal to it, except that the surface is
closed and, in the interesting case of the torus, it is also non
simply connected. As we said the case of the torus is interesting
because it is possible there to define a conserved momentum operator,
with standard commutation relations \cite{ILL}.
Therefore the mean field theory
which is obtained in the limit $e\to 0$ must still be translational
invariant in the usual sense. This is puzzling at first \cite{CWWH},
 because it is
known that the Landau levels problem is not {\it truly}
translationally invariant (only non commuting "magnetic" translations
can be defined, see for instance \cite{Fu}).
Thus, one could guess that genuine translational
invariance is only restored after corrections to the mean field
approximation are taken into account.

But in our case this is not so. In the limit $e\to 0$ the mean
field picture becomes exact and therefore it must be genuinely
translational invariant {\it per se}. And in fact translational
invariance is recovered by taking also into account the topological
degrees of freedom of the gauge field, which also do not decouple in
the limit $e\to 0$ . Thus, this limit gives a solvable theory
where also the topology of the surface plays an important role.
In fact, even before taking the $e\to 0$ limit, the Hamiltonian and
momentum contain also the contribution of the topological degrees of
freedom. In terms of the previously introduced variables $a_i$ and their
conjugate $b_i=-i\pa / \pa a_i$, the Hamiltonian gets a term proportional
to
$$
({k\over 4\pi}a_j+\epsilon_{ji} b_i)^2
$$
and the Momentum a term
$$
{1\over 2}Na_j-{2\pi N\over k}\epsilon_{ji} b_i
$$
where $N$ is the (conserved) total number of Fermions. The presence of
these terms insures that
$$
[H,P_i]=0~~~~~~[P_i,P_j]=0
$$
$H$ and $P_i$ being the total Hamiltonian and both components of the total
Momentum respectively. This structure survives the $e\to 0$ limit.

This limiting theory contains two energy scales (forgetting the
$\tilde A_i$ field which decouples). Namely  the standard energy
difference between Landau levels
$${\cal E}={e|F^b|\over M}={2\pi \rho_0 \over kM}$$
where $M$ is the particles' mass, and the energy excitation of the
topological degrees of freedom which turns out to be equal to the sum
${\cal E}+\mu$.

We will mostly discuss the case where the two scales
are of the same order of magnitude, i.e. $\mu =s{\cal E}$ with $s$
finite.
In the case $\mu >> {\cal E}$ the topological degrees of freedom
are frozen in their ground state, which in this respect is the situation
of the pure CS theory (corresponding to the opposite limit
$e\to \infty$). Thus, to be precise, the Anyon mean field theory
corresponds to the limit $e\to 0$ of the Maxwell-CS theory when
$\mu >> {\cal E}$. But actually
the ground states (at fixed total
momenta) for $\mu \sim {\cal E}$ are similar to the ones for
$\mu >> {\cal E}$ and in fact identical for the part referring to the
particles, the energy eigenvalues differing by a finite
rescaling. Thus, it is convenient to discuss the general case.

Notice that in the limit $e\to 0$ at fixed $\mu$ the CS parameter $k$
goes to infinity. Thus, it would seem that the Anyons, whose mean field
theory we have obtained, are actually ordinary Fermions since as we said
their statistical angle is $\theta =\pi (1-1/k)\to \pi$. However, a closer
inspection indicates that the case of ordinary Fermions corresponds to
$k\sim N$ and therefore to a vanishing, or "microscopic", gap:
${\cal E}\to 0$ in the thermodynamical limit. In our case instead we keep
${\cal E}$ finite. This implies that the density of particles is high
enough that even in the limit $k\to \infty$ the correction to the
statistical angle $\Delta \theta =-\pi /k$ provides a relevant effect.

We are interested in particular in the ground state of our
$N$-body system, which we find to be "protected" against
perturbations by the energy gap $\Delta E \geq {\cal E} $. Since
we have also total momentum operators $P_x,P_y$ which commute among
themselves and with the Hamiltonian, we can look for simultaneous
eigenstates of the total momentum and the Hamiltonian, describing the flow
of the particles along the torus. We find
persistent currents, i.e. we find that for some quantized eigenvalues
of the total momentum's components the motion is protected against
disturbances. The corresponding eigenstates can be interpreted as
describing a sort of motion
of a condensate of $N/k$ groups of $k$ particles,
where all the groups have the same momentum. This is seen by the value of
these exceptional eigenvalues of the total momentum, which are found to be
$P_i={N\over k}n_i$ where $n_i$ are integers (the momentum is naturally
quantized in units of the inverse of the fundamental lengths of the torus,
here put equal to 1).
Other eigenvalues of the
total momentum are also possible i.e. generically $P_i=n_i$,
but then the energy gets an
additional contribution equal to the above said gap.
Therefore a small perturbation corresponding
to the scattering of a few particles and
thus to a slight alteration of the total momentum is not possible
if an energy of the order of the gap is not available. For charged
particles those states describe therefore superconducting currents.

Also, one can compute the particles' density in these eigenstates
and one finds a constant value. This is a noncompletely trivial result,
related to the implementation of translational invariance: in fact the
expectation value of the density is obtained by integrating the modulus
square of the wave function over all the particles' positions except one
and $also$ over the topological components of the gauge field. Thus, the
mean field theory we have obtained is indeed $selfconsistent$, since the
constant background magnetic field corresponds to a constant density.

It is interesting to note that the above features, which we tried to
summarize by saying that they describe a condensate, are obtained in a
first quantization formalism, and the nontrivial manybody wave function is
explicitly known. This is probably quite unusual. Let us stress that the
above results are exact in the limit $e \to 0$.

We believe that the exact results described above provide a zero order
scenario of the properties of a fluid of Anyons, and moreover they
can also be used as a
starting point of a sistematic expansion (in powers of $e$) for taking into
account the effects of the deformation from the selfconsistent uniform
density case described above.


\begin{thebibliography}{9}
\def\bi{\bibitem}

\bi{MEAN} R.Iengo and K.Lechner, Nucl. Phys. {\bf B384} (92) 541.
\bi{Rep}R.Iengo and K.Lechner, "Anyon Quantum Mechanics and Chern-Simons
        theory"  Phys.Rep. {\bf 213} (92) 179.
\bi{BEMS}A.P. Balachandran, E. Ercolessi, G. Morandi and A.M.Srivastava,
         "Hubbard Model and Anyon Superconductivity", World Scientific
          (1990).
\bi{DJT}S. Deser, R. Jackiw and S. Templeton, Phys. Rev. Lett. {\bf 48}
        (82) 975; Ann. Phys. (NY) {\bf 140} (82) 372.
\bi{SS} G.W. Semenoff and P. Sodano, Nucl. Phys. {\bf B328} (89) 753.
\bi{Po1}A.P. Polychronakos, Ann. Phys. {\bf 203} (90) 231.
\bi{ILL}R. Iengo, K. Lechner and Dingping Li, Phys. Lett. {\bf B269}
        (91) 109.
\bi{WF} {\it Fractional Statistics and Anyon Superconductivity}, edited
        by F. Wilczek, World Scientific,
        Singapore--New Jersey--London--Hong Kong (1990).
\bi{Ho1}Y. Hosotani, Phys. Rev. Lett. {\bf 62} (89) 2785.
\bi{Po3}A.P. Polychronakos, Phys. Lett. {\bf B241} (90) 37.
\bi{IL1}R. Iengo and K. Lechner, Nucl. Phys. {\bf B346 [FS]} (90) 551.
\bi{IL2}R. Iengo and K. Lechner, Nucl. Phys. {\bf B364 [FS]} (91) 551.
\bi{La4}R.B. Laughlin, Phys. Rev. Lett. {\bf 60} (88) 2677.
\bi{FHL1}A.L. Fetter, C.B. Hanna and R.B. Laughlin, Phys. Rev. {\bf B39} (89)
        9679.
\bi{FHL2}C.B. Hanna, R.B. Laughlin and A.L. Fetter, Phys. Rev. {\bf B40} (89)
        8745.
\bi{HC}Y. Hosotani and S. Chakravarty, Phys. Rev. {\bf B42} (90) 342.
\bi{CWWH}Y.H. Chen, F. Wilczek, E. Witten and B.I. Halperin,
         Int. J. Mod. Phys. {\bf B3} (89) 1001.
\bi{Fu} S. Fubini, Int. J. Mod. Phys. {\bf A5} (90) 3533.


\end{thebibliography}
\end{document}